\documentclass[a4paper,11pt]{article}
\pdfoutput=1 

\usepackage{jheppub} 

\usepackage[T1]{fontenc} 

\title{\boldmath Physics in a general length space-time geometry:  Call for experimental revision of the light speed 
anisotropy}

\author{Qasem Exirifard,} 
\affiliation{School of Physics,\\Institute for Research in Fundamental Sciences (IPM),\\ Tehran,  Iran}
\emailAdd{exir@theory.ipm.ac.ir}
\preprint{IPM/P-2013/004}

\abstract{We present a phenomenological model for the nature in the Finsler and Randers space-time geometries. We show that  the parity-odd light speed anisotropy perpendicular to the gravitational equipotential surfaces  encodes the deviation from the Riemann geometry toward the  Randers geometry. We utilize an asymmetrical ring resonator and  propose a setup in order  to directly measure this deviation. We  address the constraints that  the current technology will impose on the deviation should the anisotropy be measured  on the Earth surface and the orbits of artificial satellites. }

\begin{document} 
\maketitle
\flushbottom

\section{Introduction}
Consider the word-line  of a particle  parametrized by $\tau$. Physics adheres to the Riemann geometry from outset and  defines the length of a curve between two points of $p$ and $q$  by  
\begin{subequations}
\label{metricRiemann}
\begin{eqnarray}
ds^2 &=& g_{\mu\nu} dx^\mu dx^\nu = g_{\mu\nu} \dot{x}^\mu \dot{x}^\nu (d\tau)^2\,, \\
L_{\| p-q\|} &=& \int_q^p  d\tau \sqrt{ g_{\mu\nu} \dot{x}^\mu \dot{x}^\nu}\,.
\end{eqnarray}
\end{subequations}
This is due to the metric postulates  which in turn are based on the Einstein Equivalence Principles. The EEP can be summarized in:
\begin{itemize}
\item WEP: Weak Equivalence Principle. 
\item LLI: Local Lorentz Invariance.
\item LPI: Local Position Invariance.
\item NIMO: No Intrinsic Memories for Objects.
\end{itemize}
The last item, most often assumed tacitly  in EEP,  means  that the history of an object does not effects its physical properties. For example two identical atomic clocks at the same point and being inertial with respect to each other - are assumed to- work at the same rate regardless of where they had been, or what velocity they had in the past.   Perhaps we should provide test models in order to test each of the EEP assumptions. Generalized geometries provide a framework accommodating   some of the test models.

We notice that yhe definition of length given in \eqref{metricRiemann} is not the the only mathematical possibility. Length can be defined in the Finsler geometry:
\begin{equation} 
\label{finsler}
L_{\| p-q\|}  = \int_q^p  d\tau L(x^\mu, \dot{x}^\mu)\,,
\end{equation}
where $L(x,\dot{x})$ is the functional of the finsler geometry and it satisfies:
\begin{equation}
L(x, \lambda \dot{x})= \lambda L(x, \lambda \dot{x}) ,\quad \lambda >0\,,
\end{equation}
Finsler geometry respects WEP, LPI and NIMO conditions. LLI in general is broken by the Finsler geometry.  So the Finsler geometry also provides a test model for LLI. For
\begin{equation}
L(x^\mu, \dot{x}^\mu) = \sqrt{ g_{\mu\nu} \dot{x}^\mu \dot{x}^\nu}\,,
\end{equation}
the Finsler geometry reduces to the Riemann geometry.  Connection, covariant derivatives and tensors that carry geometrical properties of the Finsler geometries are known. 
The action of gravity  \cite{Pfeifer:2011xi} and electrodynamics \cite{Pfeifer:2011tk}  are known in the Finsler manifolds too. 

We also can define the length by:
\begin{subequations}
\label{nonlocalgeometry}
\begin{eqnarray}
L_{\| p-q\|}  &=& \int_q^p  d\tau L(A^\mu(\tau), B^\mu(\tau))\,,\\
A^\mu(\tau) & =& \int d\tau' k_1(\tau-\tau') x^\mu(\tau') \,,\\
B^\mu(\tau) & =& \int d\tau' k_2(\tau-\tau') \dot{x}^\mu(\tau')\,,
\end{eqnarray}
\end{subequations}
where  $\tau$ is the proper time, and $k_1$ and $k_2$ are functions encoding how much memory an intrinsic particle can have. 
The nonlocal geometry of \eqref{nonlocalgeometry}  respects  LPI but breaks ``WEP'', LLI and NIMO. 
 For
\begin{equation}
k_1(\tau-\tau')=k_2(\tau-\tau') = \delta(\tau-\tau')\,,
\end{equation}
 \eqref{nonlocalgeometry} 
 reduces to the Finsler geometry in case that 
 \begin{equation}
 L(A^\mu(\tau),\lambda B^\mu(\tau)) \,=\, \lambda L(A^\mu(\tau), B^\mu(\tau)),~~\forall \lambda>0\,.
 \end{equation}
 The nonlocal-nonlinear connection, the covariant  derivative and the analog of the Riemann tensor can be defined 
 for the nonlocal geometry of  \eqref{nonlocalgeometry} \cite{NLexir}. 

No doubt that we have preferred  the Riemann definition \eqref{metricRiemann} over \eqref{finsler} and \eqref{nonlocalgeometry} . But why should the nature choose the Riemann definition of length over other possibilities, even in the scales that we have not yet directly explored? A glimpse to the history of science illustrates that the nature has proven not to infinitely follow our leads. Furthermore notice that our  obsession with the Riemann geometry  has resulted in the problem of quantum gravity (serious difficulties in quantizing the Riemann-Cartan geometry), and the failure in describing the cosmo in its large scale without modifying either the standard model of particle physics (dark matter /energy paradigm \cite{CDM}) or the Einstein-Hilbert general relativity (MOND/MOG paradigm \cite{MOND}). There is no mathematical proof that generalizing geometry can alleviate  the quantum gravity's problem. However since some of  the generalized theory breaks LLI one expects to resolve the quantum gravity's problem by extended geometries.  The dark matter  and energy   problems, on the other hand,  can be resolved in the Finsler geometry   \cite{Chang:2008yv, Chang:2009pa, Li:2012qga}. Finsler geometry also accommodates the spontaneous Lorentz invariance violation \cite{Kostelecky:2011qz}. Some of the observational effects of a class of the Finsler geometries are explored in   \cite{Lammerzahl:2012kw}. Connection between Finsler-Randers space-time and DGP gravity model is explored in  \cite{DGP}

This article asks how much precise  the space-time can be approximated to the Riemann geometry. Perhaps ``very good'' is a reply. This answer, however, is qualitative rather than being  quantitative.  In order to provide the quantitative answer,  we create  a phenomenological model for the nature in Finsler geometry within section \ref{phenomenologicalmodel} . This allows us to systematically study the possible deviation of the space-time geometry from the Riemann geometry. In section \ref{lengthEarth} we derive the space-time geometry near the Earth in the model presented in the  section \ref{phenomenologicalmodel}.  In section \ref{LSanisotropyEarth} we study local light speed anisotropy near the Earth. We show that the parity-odd light speed anisotropy perpendicular to the gravitational equipotential surfaces encodes the deviation from the Riemann geometry toward the Randers/Finsler geometry. In order to measure this anisotropy, following \cite{Trimmer:1973nn}, in section \ref{LImatter} we first present the light speed anisotropy in a medium. We then address  the precision for measuring the parity-odd light speed anisotropy parameters that is and can be achieved by the current technology. Section   \ref{Precisions} argues that  the current technology allows reaching the precision of $p_{\mbox{\tiny exp}}=\frac{\delta c}{c}=10^{-13}$, and improving the technologies allows reaching $p_{\mbox{\tiny exp}}=10^{-17}$ and beyond.  We note that a set of the deviations from the Riemann geometry are better constrained in stronger gravitational fields  while another set are better constrained in weaker gravitational fields.  We, therefore, in section \ref{Proposals} propose to measure the parity-odd light speed anisotropy perpendicular to the equipotential surfaces on the Earth surface and also in the orbits of artificial satellites.  We derive the allowed range of the deviation parameters should the anisotropy be measured on the Earth surface and on the geosynchronous orbit with the precision of $p_{\mbox{\tiny exp}}$ and no signal be observed. In section \ref{Iran} we also discuss with what precision these deviations can be measured in Iran, on the ground and within  the orbits of the Iranian satellites.

 
\section{A Phenomenological Model}
\label{phenomenologicalmodel}
We consider Finsler geometries where the length element is defined by 
\begin{eqnarray}\label{lgeneral}
dl &= & \sum_{n=1}^{\infty}  (a_{\mu_1 \cdots \mu_n}(x) dx^{\mu_1}\cdots dx^{\mu_n})^{1/n}  \\
 &=& (a_{\mu_1 \mu_2} dx^{\mu_1} dx^{\mu_2})^{\frac{1}{2}} + a_{\mu_1} dx^{\mu_1} + (a_{\mu_1 \mu_2 \mu_3} dx^{\mu_1} dx^{\mu_2} dx^{\mu_3})^{\frac{1}{3}}+ \cdots\,.
\end{eqnarray} 
Since our space-time geometry can be approximated with the Riemann geometry then $n=2$ is  the largest term in \eqref{lgeneral}. We consider $n=2$ as the leading term is the definition of the length.  The rest of the terms ($n\neq 2$) should be considered as the perturbations (sub-leading terms) to the length element.  In this article we truncate the perturbative series to the simplest perturbation ($n=1$) and we consider:
\begin{equation}
dl =  (g_{\mu\nu} dx^\mu dx^\nu)^{\frac{1}{2}} + a_{\mu} dx^\mu\,.
\end{equation}
This is the Randers space-time  geometry \cite{Randers}. So to define the length element, we have endowed the  manifold of the space-time geometry with a symmetric tensor $g_{\mu\nu}$ and a vector field $a_\mu$:
\begin{eqnarray}
x^\mu & \to & \tilde{x}^\mu\,,\\
g_{\mu\nu} &\to & \tilde{g}_{\mu\nu} = \frac{\partial x^\alpha }{\partial \tilde{x}^\mu }  \frac{\partial x^\beta }{\partial \tilde{x}^\nu} g_{\alpha\beta}\,,\\
a_\mu &\to & \tilde{a}_\mu = \frac{\partial x^\alpha }{\partial \tilde{x}^\mu }  a_\alpha\,.
\end{eqnarray}
 Note that the Randers/Finsler geometry leave the general covariance  intact but break the Lorentz symmetry: the vacuum expectation value of $a_\mu$ breaks the Lorentz symmetry.  We refer to  $g_{\mu\nu}$ as the metric. We, however, notice that the metric alone does not define the length element.  We then utilize the following perturbative series:
\begin{eqnarray}
g_{\mu\nu} = g_{\mu\nu}^{(0)}+ \epsilon g_{\mu\nu}^{(1)}\,,\\
a_\mu = a_\mu^{(0)} + \epsilon a_\mu^{(1)}\,,
\end{eqnarray}
where $\epsilon$ presents the systematical parameter of the perturbation. We keep $\epsilon$ to track the perturbation but at the end of the  calculation  we set $\epsilon=1$. The perturbation for the length element then follows:
 \begin{eqnarray}
dl^{(0)} & =& (g_{\mu\nu}^{(0)} dx^\mu dx^\nu)^{\frac{1}{2}}+ a_\mu^{(0)}  dx^\mu \,,  \\
\label{dl}
dl= dl^{(0)} + \epsilon dl^{(1)} &=&  \left((g_{\mu\nu}^{(0)} + \epsilon g_{\mu\nu}^{(1)}) dx^\mu dx^\nu\right)^{\frac{1}{2}} +\epsilon a_{\mu}^{(1)}  dx^{\mu}\,,
\end{eqnarray}
which will be studied by the following algorithm: 
\begin{enumerate}
\item The leading order geometry ($dl^0$) is given to us: $g_{\mu\nu}^{(0)}$ is known.  The  leading order  has a vanishing vector field: 
\begin{equation}
a_\mu^{(0)}=0.
\end{equation}
\item  As any perturbative theory,  $a_\mu^{(1)}$ can be expressed in the term of the fields present in the  leading approximation.   We assume that $a_\mu^{(1)}$ can be expressed   in term of $g_{\mu\nu}^{(0)}$ and its derivatives.
\item We study local light speed anisotropy  produced by $a_\mu^{(1)}$.  Note that no corrections added to $g_{\mu\nu}^{(0)}$ produces local light speed anisotropy. So within our study we need not know  $g_{\mu\nu}^{(1)}$.  
\end{enumerate}
One and 2 requires that $a_\mu^{(1)}$ must be a functional of the  Riemann tensor and its covariant derivatives constructed out from the metric $g_{\mu\nu}^{(0)}$:
\begin{equation}
a_\mu^{(1)} \,=\, f_\mu( R_{\mu\nu\lambda\eta}, \nabla_\mu,  g_{\mu\nu}^{(0)})\,,
\end{equation}
where $f_\mu$ must have an expansion in term of the vector invariants of the metric $g_{\mu\nu}^{(0)}$. All the vector invariants can be written in term of the covariant derivatives of the scalar invariants. So
\begin{equation}
f_\mu = \nabla_\mu S( R_{\mu\nu\lambda\eta}, \nabla_\mu,  g_{\mu\nu}^{(0)})\,,
\end{equation}
where $S$ is a scalar constructed out from the Riemann tensor and its covariant derivatives (which in turn are constructed out from  the metric $g_{\mu\nu}^{(0)}$).
 We notice that  the $g_{\mu\nu}^{(0)}$ representing the metric in vacuum (as well as light density of matter) holds 
\begin{eqnarray}
R = R_{\mu\nu} = 0\,, \\
R_{\mu\nu\lambda\eta} R^{\mu\nu\lambda\eta} \neq 0 \,,
\end{eqnarray} 
where $R$ and $R_{\mu\nu}$ are  the Ricci scalar and tensor. So the first  non-trivial invariant scalar around the Earth space-time geometry is  $R_{\mu\nu\lambda\eta} R^{\mu\nu\lambda\eta} $.  We confine our phenomenological model to the cases where  the following expansion series  holds:
\begin{equation}
\label{amu}
a_\mu^{(1)} \,\equiv\,\sum_{m} c_m \nabla_\mu \left(R_{\mu\nu\lambda\eta} R^{\mu\nu\lambda\eta}\right)^m \,,
\end{equation}
where $c_m$ are constant parameters chosen by the nature. Note that we do not restrict $m$ to integer numbers and $[c_m]= L^{4m+1}$. 

The standard theory of general relativity realized in the Riemann geometry from outset assumes  that 
\begin{equation}
c_m = 0 \quad \forall m\,.
\end{equation}
We should, however, ask with what precision we know that $c_m=0$ for each $m$. We should measure $c_m$ with some precision. 

 
 \section{Length element near the Earth}
 \label{lengthEarth}
 We approximate the leading term describing the space-time metric around the Earth  to   the Schwarzschild solution:
 \begin{eqnarray}\label{g00Schwarzschild}
  g_{\mu\nu}^{(0)}dx^\mu dx^\nu\,=\,(1-\frac{2 G M_\oplus}{c^2 r}) dt^2 + \frac{dr^2 }{1-\frac{2 G M_\oplus}{c^2 r}} + r^2 d\Omega^2 \,,
 \end{eqnarray}
 where $M_\oplus$ is the mass of the Earth and $r$ is the distance from the center of the Earth:
 \begin{equation}
 M_\oplus\,=\, 5.97 \times 10^{24}~\mbox{ kilograms}\,,
 \end{equation}
 and $G$ is the Newton's gravitational constant. 
 Note that  $g_{\mu\nu}^{(0)}$ holds:
 \begin{eqnarray}
 R_{\mu\nu\lambda\eta} R^{\mu\nu\lambda\eta}& =& \frac{48 G^2 M_\oplus^2}{c^4 r^6}\,, \\
 a_\mu^{(1)}=\sum_{m} c_m \nabla_\mu(R_{\mu\nu\lambda\eta} R^{\mu\nu\lambda\eta})^m& =& -  \sum_{m}   6m c_m\left(\frac{48 G^2 M_\oplus^2}{c^4} \right)^m r^{-6m-1}  ~\delta^r_{~\mu}\,.
 \end{eqnarray}
 where $R_{\mu\nu\lambda\eta}$ is the Riemann tensor constructed out from   $g_{\mu\nu}^{(0)}$.
 The length element near the Earth then follows from \eqref{dl} and \eqref{amu}:
 \begin{equation}
 dl= (g_{\mu\nu}^{(0)} dx^\mu dx^\nu+\epsilon g_{\mu\nu}^{(1)} dx^\mu dx^\nu)^{\frac{1}{2}} + \epsilon \tilde{c}(r) dr\,,
 \end{equation}
 where 
 \begin{equation}\label{tildeccm}
 \tilde{c}(r) \,\equiv \, -\sum_{m} 6 m c_m \left(\frac{48 G^2 M_\oplus^2}{c^4} \right)^m r^{-6m-1} \,.
 \end{equation}
 Perhaps both $g_{\mu\nu}^{(1)}$ and $\tilde{c}$ affect the orbits of satellites but  $g_{\mu\nu}^{(1)}$ does not affect the local light speed anisotropy. Some of the observable effects in a class of spherically symmetric static Finsler geometries  are studied in \cite{Lammerzahl:2012kw}. We study a class of observables which is not studied in \cite{Lammerzahl:2012kw}. The observables we study are only sensitive to $\tilde{c}(r)$. Those studied in \cite{Lammerzahl:2012kw} are sensitive to a combination of $\tilde{c}$ and $g_{\mu\nu}^{(1)}$.


 \section{Light speed anisotropy in the vacuum near the Earth}
 \label{LSanisotropyEarth}
  In the study of local light speed anisotropy  it is better to  choose the local Riemann coordinate such that $\eta_{\mu\nu}$ replaces $g_{\mu\nu}^{(0)}+\epsilon g_{\mu\nu}^{(1)}$. This simplifies the length element to:
 \begin{equation}\label{dleta0}
 dl= (\eta_{\mu\nu} dx^\mu dx^\nu)^{\frac{1}{2}} +  \epsilon \tilde{c}(r) dr + O(\epsilon^2)\,,
  \end{equation}
 where $\hat{r}$ is the radial coordinate of the standard coordinate of the Earth and $r$ is the distance to the center of the Earth. Now we set $\epsilon=1$:   
\begin{eqnarray}\label{dleta}
 dl&=& (\eta_{\mu\nu} dx^\mu dx^\nu)^{\frac{1}{2}} +  \tilde{c}(r) dr\, \\
 &=& (-dt^2+ \sum_{i}(dx^i)^2)^{\frac{1}{2}} +  \tilde{c}(r) dr\, .
  \end{eqnarray}
 Ref. \cite{Pfeifer:2011tk}  provides a definition of the Finsler geometries that does not include Randers geometry. It then   defines electrodynamics such that  null geodesics are photon's world lines. Following  \cite{Pfeifer:2011tk},  we simply assume that null geodesics represent world lines of photons.  Let the world line of a photon be parametrized by $\tau$: $x^\mu = (c t(\tau), x^i(\tau))$. Then \eqref{dleta} yields
 \begin{equation}\label{41ignore}
 - c^2 \dot{t}^2 + \dot{\vec{x}}^2 = \tilde{c}^2(r) \dot{r}^2\,,
 \end{equation} 
 where ${}^. \equiv \frac{d}{d\tau}$. We shall ignore terms which are quadratic in $\tilde{c}(r)$ (since they are at the order of $\epsilon^2$).  Eq. \eqref{41ignore} then results:
 \begin{equation}
 |\frac{d\vec{x}}{dt}| = c\,,
 \end{equation}
 Notice that $|\frac{d\vec{x}}{dt}|$ is not the physical light speed. The light speed by definition is the length that light travels per unit of time. We first choose the standard observer. We then employ the Edington synchronization method to synchronies the clocks with the clock of  the standard observer(the clock attached to the standard observer).  The standard observer has a preferred time: his/her proper time.   We represent this choice of time by $t$. We consider the hyper surfaces of $t= cte$ in order to define the length element  in space. Recalling the length element in the space-time then induces the definition of the length in the space:
 \begin{equation}
\mbox{Eq. \eqref{dleta} and}~ (dt=0) ~\to~ ||dx|| =  |dx| +  \tilde{c}(r) dr\,,
 \end{equation} 
 where $|dx|$ is the Euclidean length:  $|dx|=\sqrt{\sum_i (dx^i)^2}$.  A light pulse takes $dt$ to travel $d\vec{x}$.  So the light speed follows:
 \begin{eqnarray}
  c(\hat{n}) \,=\, \frac{||dx||}{dt}& = & c ( 1 + \tilde{c}(r) \hat{n}.\hat{r}) \label{cn}\,,
 \end{eqnarray}
 where 
 $\hat{n}$ 
 is the light propagation direction, $\hat{r}$ is the radial direction, $c(\hat{n})$ is the light speed in direction of $\hat{n}$.
 
 Eq. \eqref{cn} presents a parity-odd anisotropy for the light speed: 
 \begin{equation}
 c(\hat{n})+c(-\hat{n}) = 2 c\,,
 \end{equation}
 the size of  which is better expressed by
 \begin{equation}
 \label{dcc1}
 \frac{\delta c}{c} = 2 |\tilde{c}(r)|\,.
 \end{equation} 
 In other words \eqref{cn}  represents a one-way light speed anisotropy. So the parity-odd light speed anisotropy in the direction of $r$, the direction  perpendicular to the gravitational equipotential surfaces encodes the deviation from the Riemann geometry towards the Randers geometry.

 
 \section{Light propagation in a  medium}
  \label{LImatter} 
 Let it be highlighted that   one of the  earliest model for the light speed anisotropy is Robertson-Mansouri-Sexl model \cite{Robertson,Mansouri}.    The model assumes that the maximum attainable velocity (the Einstein speed) in every direction coincides to the light speed in that direction. The model neither addresses what the essence of light is nor  considers the light interaction with bulk of matter. Its only inputs   are the light pulses  in the vacuum. The model argues  that the parity odd light speed anisotropy is the artifact of the clock synchronization method albeit in the circumstances that one only considers  the light pulses in the vacuum and  assumes that the Einstein speed in each direction coincides to the light speed in that direction. 
 
 A comprehensive theory of the nature, however, should also address how light interacts with matter.  Once this interaction is given, the parity-odd light speed anisotropy in the vacuum  induces a light speed anisotropy in the matter medium, a quantity that is measurable without the need of the clock synchronization procedure, in fact the Trimmer interferometer measures the induced anisotropy  \cite{Trimmer:1973nn}.    In order to describe the propagation of light in matter within a given arbitrary space-time geometry, we accept the hypothesis of \cite{Trimmer:1973nn}. We assume that the light speed in a direction in a homogeneous isotropic matter is equal to the light speed in the vacuum in the same direction divided  by the  index of refraction of the matter:
  \begin{equation}
  c_{\mbox{\tiny matter}}(\hat{n}) = \frac{1}{n}c(\hat{n})\,,
  \end{equation}
  where as before, $\hat{n}$ is the direction of the light propagation, $c_{\mbox{\tiny matter}}(\hat{n})$ represents the light speed in matter, and $n$ is the  index of refraction the medium.\footnote{The Standard-Model-Extension (SME) model  \cite{Kostelecky:2008ts} in nutshell assigns  the anisotropy of the vacuum to the permeability of the vacuum. It then assumes that  the permeability of matter adds to the permeability of the vacuum. This gives the interaction between matter and light in the SME. The SME is a global model for the light speed anisotropy. The model presented here is a local one. It would be interesting to provide a local extension of the SME.} This assumption enables us to measure the parity-odd light speed anisotropy with the Trimmer setup  \cite{Trimmer:1973nn}, or with its modern generalizations \cite{Exirifard:2010xp,Baynes:2011nw}.  In other words, adding the Trimmer hypothesis makes our model a test model for the local light speed anisotropy.

\section{Precisions achieved  by the current technology}
\label{Precisions}
\begin{figure}[tbp]
\centering 
\includegraphics[width=.45\textwidth]{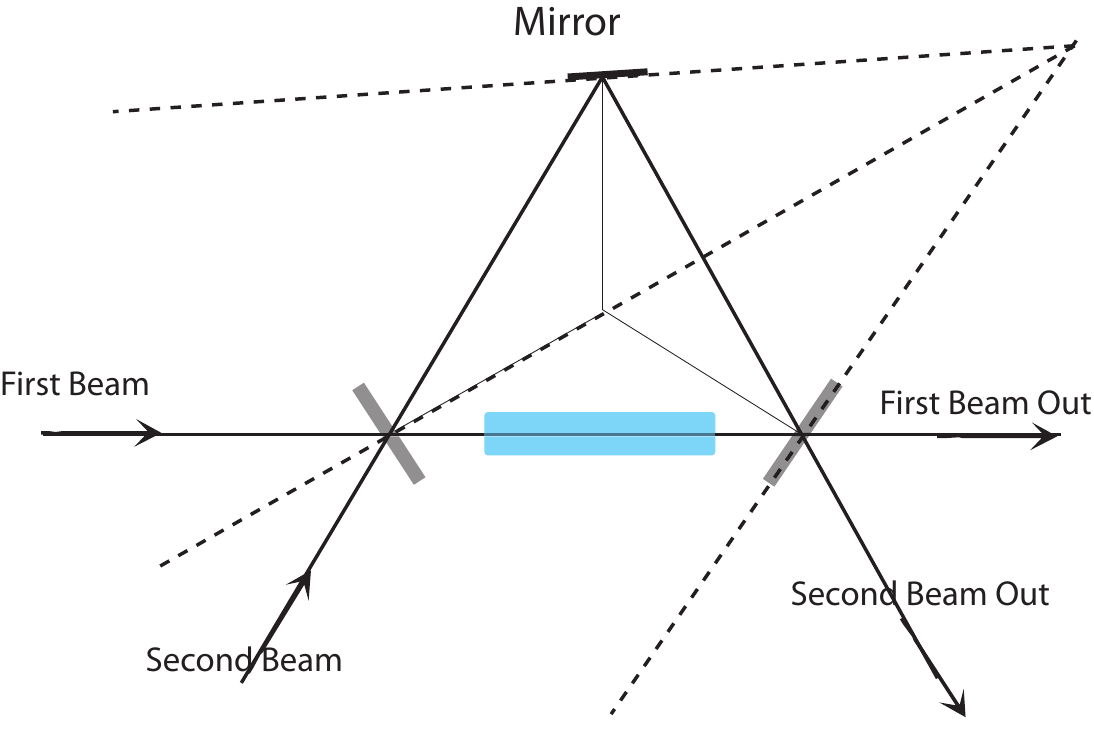}
\includegraphics[width=.45\textwidth]{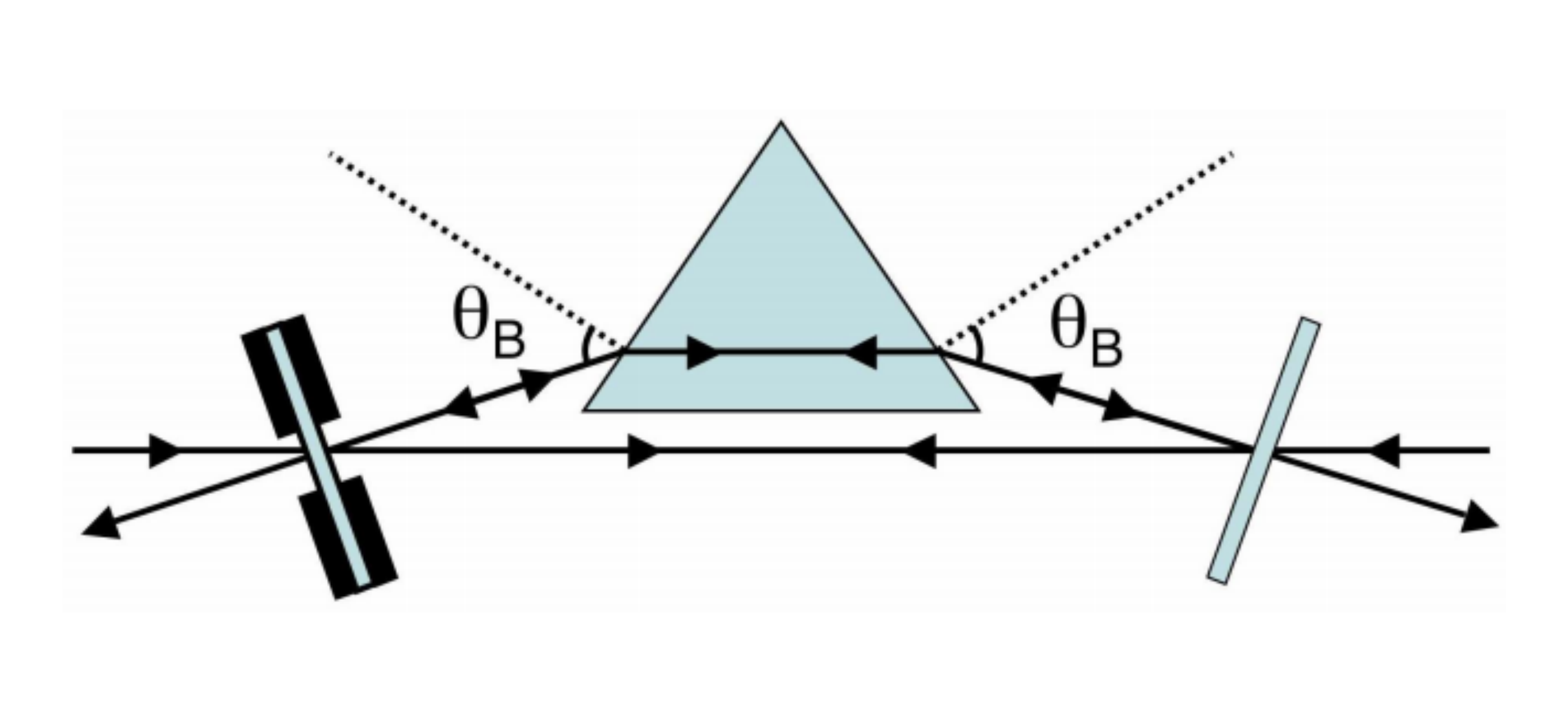}
\caption{\label{fig:1} Left hand: The proposed asymmetrical triangular resonator of ref.  \cite{Exirifard:2010xp} . Right hand: The asymmetric ring resonator with a prism
at the Brewster's angle $\theta_B$ of the performed experiment by ref.  \cite{Baynes:2011nw}. }
\end{figure}

Ref \cite{Eisele:2009zz,Muller:2007zz} uses modern optical resonators on a table rotating parallel to the gravitational equipotential surface and reports that the parity even light speed anisotropy parameters should be smaller than one part in $10^{-17}$.  Ref.  \cite{Baynes:2011nw} uses a modification of the setup presented in \cite{Exirifard:2010xp} on a fixed table and reports that the isotropic Lorentz violating parameter of SME model  is less than $10^{-9}$ while the  parity-odd parameters of the SME are smaller than $10^{-13}$, fig. \ref{fig:1}.

We do assume that the precision of  \cite{Baynes:2011nw}  can be achieved for measuring the parity odd light speed anisotropy perpendicular to the gravitational equipotential. In other words, we assume that 
\begin{enumerate}
\item An optical table is designed such that it rotates perpendicular to the gravitational equipotential surface. 
\item The experiment of \cite{Baynes:2011nw} is repeated on the above optical table. Note that the gravitational deformation  is automatically canceled in the setup of  \cite{Baynes:2011nw}  and \cite{Exirifard:2010xp}  because these setup measure the beat frequency of lights moving on the same path. Also note that the Sagnac signals can be removed from the data by post processing  the data should two ring resonators of different areas be used on the same rotating table.
\item The experiment of \cite{Baynes:2011nw} is repeated in the space.   
\end{enumerate}
Ref.  \cite{Baynes:2011nw}   has used microwave frequency and reached  the precision of $p_{\mbox{\tiny exp}}= 10^{-13}$. Ref. \cite{Exirifard:2010xp} argues that if  optical frequencies are used instead of the microwave frequencies then the precision of \cite{Eisele:2009zz} ($p_{\mbox{\tiny exp}}=10^{-17}$) can be achieved. So we assume that the precision of $p_{\mbox{\tiny exp}}= 10^{-13}$ can  be achieved, and the precision of $p_{\mbox{\tiny exp}}= 10^{-17}$ might be achieved in measuring the parity-odd parameters perpendicular to the gravitational equipotential surfaces, both on the ground and in the artificial satellites. 

\section{Constraints on the  deviation from the Riemann geometry}
\label{Proposals}
\begin{figure}[tbp]
\centering 
\includegraphics[width=.45\textwidth]{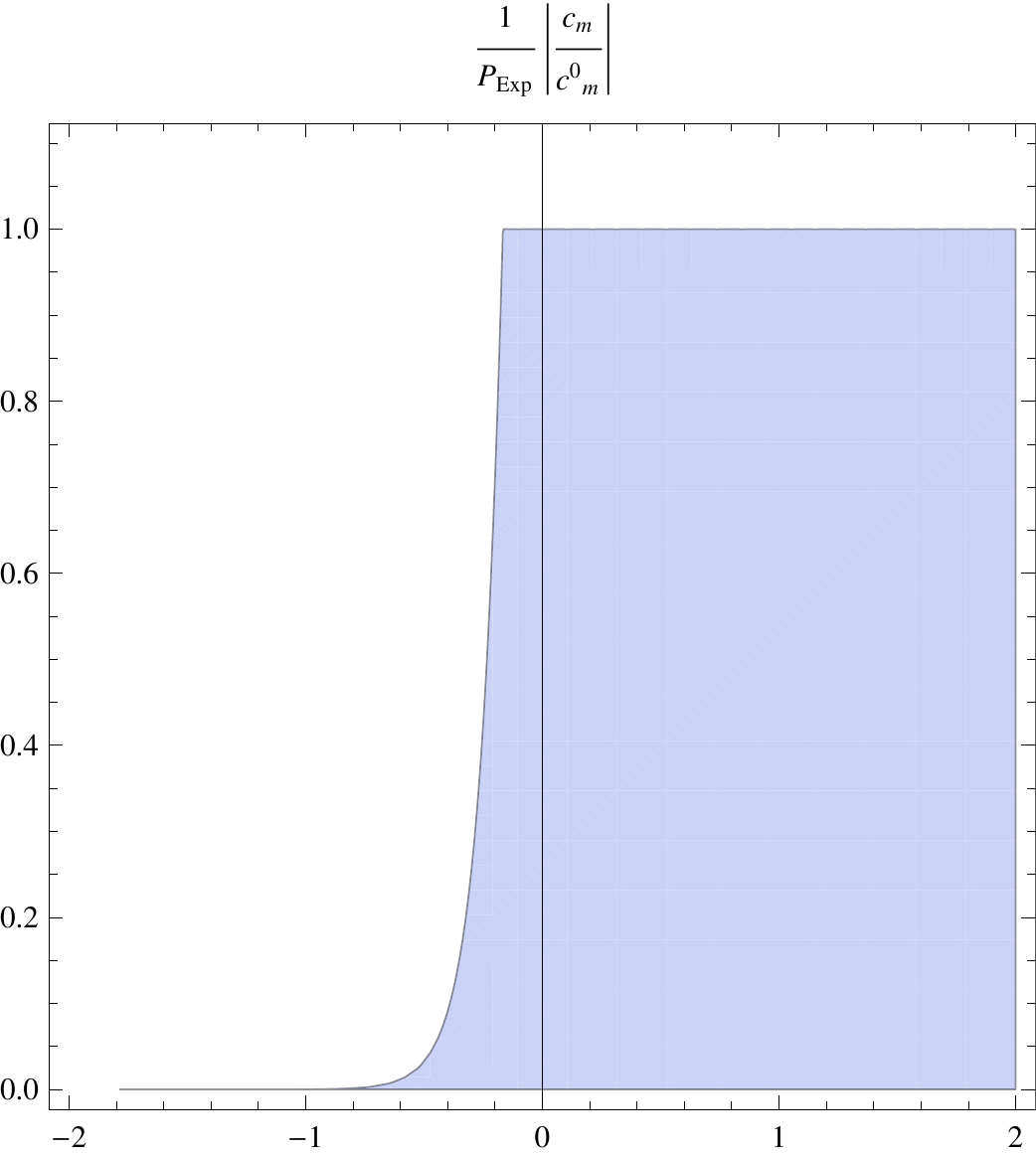}
\caption{\label{fig:2} The dashed area represents the allowed range for the deviation parameters toward the Randers/Finsler space-time geometry should the parity odd light speed anisotropy be measured on the Earth surface and the  geosynchronous orbit with the precision of $p_{\mbox{\tiny exp}}$ and no signal be observed.}
\end{figure}
Suppose that the parity-odd light anisotropy parameters perpendicular to the gravitational equipotential be measured at the distance of $r$ from the center of the Earth with the precision of $p_{\mbox{\tiny exp}}$. As discussed in the section \ref{Precisions},   $p_{\mbox{\tiny exp}}= 10^{-13}$ can  be achieved, and the precision of $p_{\mbox{\tiny exp}}= 10^{-17}$ might be achieved. There exist two possibilities:
\begin{enumerate}
\item If we observe a signal we then conclude that the space-time geometry around the Earth has deviated from the Riemann geometry. Note that no correction in the Riemann geometry produces such a signal. 
\item If we do not observe any signal then we get a constraint on the  deviations parameters of $c_m$ . 
\end{enumerate}
If we do not observe any signal we conclude that 
\begin{equation}
\left.\begin{array}{c}
\frac{\delta c}{c} < p_{\mbox{\tiny exp}} \\ \\
\mbox{Eq.}  \eqref{dcc1}
\end{array}\right\}\to ~2 |\tilde{c}(r)| <p_{\mbox{\tiny exp}} \label{crpexp}\,.
\end{equation}
Using the expansion series of \eqref{tildeccm} and assuming that  no fine tuning occurs  we conclude that each term in \eqref{tildeccm} should hold \eqref{crpexp}. This results: 
\begin{equation}\label{Q1}
 \left|\frac{c_m}{c^0_m} \right|< p_{\mbox{\tiny exp}}  \left(\frac{r}{r_\oplus}\right)^{6m+1}\,,
 \end{equation}
 where
 \begin{eqnarray}\label{cm0}
 c_m^0 &=& \frac{r_\oplus}{12\, m} \left(\frac{c^4\, r_\oplus^6}{48\, G^2 \,M_\oplus^2}\right)^{m} \,,
  \end{eqnarray}
where  $r_\oplus$ is the average radius of the Earth:
\begin{equation}
r_\oplus \,= \, 6.3675 \times 10^6 ~\mbox{meters}\,.
\end{equation} 
Eq.  \eqref{Q1} indicates that for   $6m+1>0$ performing the experiment in a stronger gravitational field (smaller $r$) leads to a better constraint on    $c_m$ while for   $6m+1<0$ performing the experiment in a weaker gravitational field   (larger $r$) leads to a better constraint on    $c_m$. So we propose to measure the parity-odd light speed anisotropy perpendicular to the gravitational equipotential surfaces on the Earth surface (the geoid) and also within the  geosynchronous orbit. Fig. \ref{fig:2} depicts the allowed range for $c_m$ should these measurements be performed and no signals be observed. Note that in fig. \ref{fig:2} , $c_m$ would be better constrained for  $m<-1/6$  due to performing the experiment in the geosynchronous orbit.

\section{On the possibility of performing  measurements  in Iran}
\label{Iran}
\begin{figure}[tbp]
\centering 
\includegraphics[width=.45\textwidth]{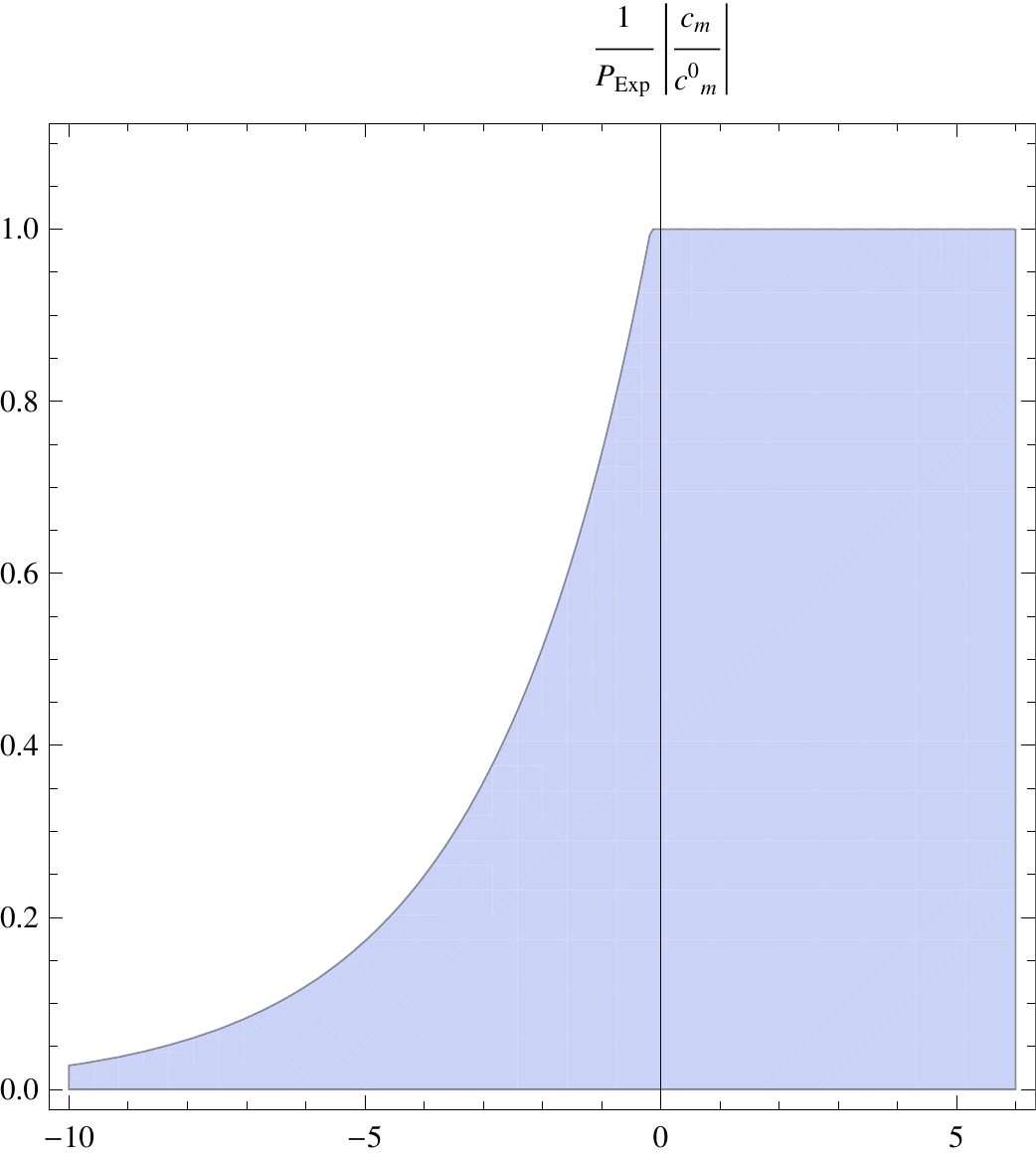}
\caption{\label{fig:3} The dashed area is the allowed range for the $c_m$ parameters  if the parity odd light speed anisotropy be measured on the Earth surface and the orbit of $400 km$ from the Earth surface with the precision of $p_{\mbox{\tiny exp}}$ and no signal be observed.}
\end{figure}

We note that various optics laboratories in Iran have the technology to measure the light speed anisotropy with the precision of about $10^{-10}$ and beyond. In other words the precision of \cite{Trimmer:1973nn} now can be achieved. We also notice that Iran Space agency has launched satellites to the low Earth orbits. The agency seems eager to perform experiments in the space, for example by sending live animals into space.  I urge my experimental colleagues in Iran, as well as the rest of the globe,  to seriously consider measuring the light speed anisotropy perpendicular to the gravitational equipotential surfaces. Any constraints on the $c_m$ parameters is welcomed.  

Fig. \ref{fig:3}  depicts the allowed range for $c_m$ should the parity-odd light anisotropy be measured on the ground and also in the orbit of $400km$ -an orbit that can be achieved by the Iranian satellites- and no signals be observed. Since there exist no  constraints on $c_m$,  the constraints in fig. \ref{fig:3} for $p_{\mbox{\tiny exp}}=10^{-10}$ will be compelling too.  

\section{Discussions}
\label{Discussions}
 In order to  review the indirect bounds on light speed anisotropy parameters in SME please refer to ref.  \cite{Exirifard:2010xm}. We note that neither of the constraints reviewed  in \cite{Exirifard:2010xm}  can be extended to the parity-odd light speed anisotropy perpendicular to the Earth surface due to the presence of the Earth. The author is not aware of any other constraint. Note that the light speed anisotropy perpendicular to the geoid has not yet been measured \cite{ModernTests}. 
 We have presented a model for the light speed anisotropy perpendicular to the gravitational equipotential and we have presented a setup to directly measure it.

We have approximated the leading order space-time geometry around the Earth to the Schwarzschild metric in the eq. \eqref{g00Schwarzschild}. Perhaps one may use a better approximation and approximate  $g^{(0)}_{\mu\nu}$ to the Kerr metric. In so doing one realizes that \eqref{amu} calculated for the Kerr metric results to a parity-odd light speed anisotropy with respect to the axis of the rotation of the Earth in directions  parallel to the gravitational equipotential surfaces. This anisotropy, however, is suppressed by the factor of $\frac{2 \pi  r_\oplus} {\mbox{\small day}\times c }=10^{-6}$.  No measurement for this suppressed parity is available but that of \cite{Trimmer:1973nn}.  Since the precision of \cite{Trimmer:1973nn} is $\frac{\delta c}{c} = 10^{-10}$, ref. \cite{Trimmer:1973nn} constrains $c_m$ parameters as if the light speed anisotropy perpendicular to the geoid were measured with the precision of $p_{\mbox{\tiny exp}}= 10^{-4}$. Considering  the precisions that can be achieved by the current technology this precision is not compelling.

We have confined our model to the cases wherein $a_\mu$ coincides to the  derivative of a scalar constructed out from the Riemann tensor, eq. \eqref{amu}. Perhaps one may consider  $a_\mu$ to be in more general forms. Also one may consider the deviation from the Riemann geometry  by considering terms of $n\neq 2$ and $n \neq 1$ in \eqref{lgeneral}.  These generalizations lead to a light speed anisotropy near the Earth due to how the Earth may affect its surrounding space-time geometry.  To constrain all possible deviations one should experimentally measure all the parity-odd and parity-even light speed anisotropy. The gravitational deformation is a big obstacle in measuring the parity-even parameters perpendicular to the gravitational equipotentials. The parity-odd parameters however can be measured. This also suggests to utilize the vacuum triangular interferometer/resonator  of ref. \cite{Trimmer:1973nn}  and \cite{Exirifard:2010xp}, and measure the higher order parity-odd coefficients of the light speed anisotropy near the Earth in all directions with respect to the Earth.

\section{Conclusions}
We have considered the possibility that the space-time geometry around the Earth deviates from the Riemann geometry towards Randers/Finsler geometry.

We have presented a model to accomadate the space-time geometry around the Earth in the Randers  geometry. The deviation from the Riemann geometry in our model is encoded in $c_m$ parameters. General Relativity realized in the Riemann geometry assumes that $c_m=0~ \forall m$. We have asked with what precision we know that $c_m$ vanish. We have shown that the $c_m$ parameters attribute to the parity-odd light speed anisotropy  perpendicular to the gravitational equipotentials due to how the Earth affects its surrounding space-time geometry.  We have proposed to utilize  the asymmetrical ring resonators on tables rotating perpendicular to the gravitational equipotential surfaces in order to directly measure $c_m$. We have addressed the precision that can be achieved by the current technology.  In order to experimentally constrain $c_m$'s for all $m$ we have suggested the experiment to be performed both on the ground and in the space. We have obtained the allowed range of the deviation parameters should the anisotropy be measured on the Earth and the geosynchronous orbit (fig. \ref{fig:2}), and also on the ground and the orbits of the Iranian satellites (fig. \ref{fig:3}).

\acknowledgments This work was supported  by the Institute for Research in Fundamental Sciences (IPM). 

\end{document}